# WaveLLDM: Design and Development of a Lightweight Latent Diffusion Model for Speech Enhancement and Restoration

Kevin Putra Santoso
*Department of Information Technology*
Institut Teknologi Sepuluh Nopember
`kevin@avalon-ai.org`

Rizka Wakhidatus Sholikah
*Department of Information Technology*
Institut Teknologi Sepuluh Nopember
`wakhidatus@its.ac.id`

R. V. Hari Ginardi
*Department of Information Technology*
Institut Teknologi Sepuluh Nopember
`hari@its.ac.id`

*Abstract*—High-quality audio is essential in a wide range of applications, including online communication, virtual assistants, and the multimedia industry. However, degradation caused by noise, compression, and transmission artifacts remains a major challenge. While diffusion models have proven effective for audio restoration, they typically require significant computational resources and struggle to handle longer missing segments. This study introduces WaveLLDM (Wave Lightweight Latent Diffusion Model), an architecture that integrates an efficient neural audio codec with latent diffusion for audio restoration and denoising. Unlike conventional approaches that operate in the time or spectral domain, WaveLLDM processes audio in a compressed latent space, reducing computational complexity while preserving reconstruction quality. Empirical evaluations on the Voicebank+DEMAND test set demonstrate that WaveLLDM achieves accurate spectral reconstruction with low Log-Spectral Distance (LSD) scores (0.48–0.60) and good adaptability to unseen data. However, it still underperforms compared to state-of-the-art methods in terms of perceptual quality and speech clarity, with WB-PESQ scores ranging from 1.62 to 1.71 and STOI scores between 0.76 and 0.78. These limitations are attributed to suboptimal architectural tuning, the absence of fine-tuning, and insufficient training duration. Nevertheless, the flexible architecture that combines a neural audio codec and latent diffusion model provides a strong foundation for future development.

## I. INTRODUCTION

Humans have always communicated, and voice plays a crucial role in daily interactions. In today's rapidly evolving digital era, audio quality has become increasingly important across various applications—from online communication and virtual assistants to text-to-speech systems, the multimedia industry, and creative content production [1], [30], [45]. High-quality audio enhances user experience in receiving information, increases audience engagement, and supports accessibility for individuals with hearing impairments [16]. However, one of the primary challenges in speech processing is quality degradation caused by environmental noise, data compression, and transmission disruptions. These issues not only degrade the user experience but also hinder accessibility and the effectiveness of voice-based services, especially under unstable network conditions or on low-power devices.

The demand for high-quality speech processing technologies continues to grow alongside the proliferation of voice-driven applications across industries. One notable use case is the restoration of damaged archival recordings due to the physical deterioration of storage media, reconstruction of missing segments caused by scratches on CDs, or compensation for packet loss in network-based voice communication [34]. Deep learning-based approaches have shown outstanding performance in speech enhancement and restoration. Numerous methods have been proposed in the literature to tackle these challenges [3], [5], [6], [12], [21], [33].

Enhancement and restoration are inverse problems [4], [6], [51]. In this context, diffusion models have demonstrated superiority over other generative models such as Generative Adversarial Networks (GANs) [5], [8], [11], [29], [43]. Probabilistic diffusion models have shown strong performance in reconstructing missing music segments of 50–300 ms, outperforming most previous state-of-the-art methods [6], [8], [9], [23], [25], [26], [34], [55]. These models exhibit a contextual understanding of sequential audio data and leverage this as a prior for sampling-based prediction [13], [14]. However, performance tends to degrade as the duration of missing segments increases. While the models can still generate the missing audio, the output is often misaligned with the broader speech context or musically inconsistent [23].

Self-supervised approaches have been employed to learn contextual speech representations through pre-training. HuBERT, for example, has been adapted as an encoder to model and manipulate latent audio representations with missing parts, while HiFi-GAN acts as a decoder that maps these representations back to the audio domain. This model performs well in reconstructing speech segments missing for 200–400 ms, although it still struggles to accurately reproduce certain vowels and consonants. Adding contextual information during processing has been shown to improve reconstruction quality over longer durations [2].

Another approach integrates multimodal information—specifically, speech audio and facial motion video—to reconstruct missing voice segments. The core model adopts a Bidirectional LSTM architecture with Connectionist Temporal Classification (CTC) decoding. Results show that performance drops significantly beyond 800 ms when relying solely on audio, whereas audio-visual approaches maintain stable performance across a range of 100–1600 ms [35].

Despite the many proposed methods, several limitations remain unresolved. First, conventional diffusion models that operate directly in the time or spectral domain (e.g., raw waveform or spectrogram) demand high computational resources and struggle with high data variance, making them suboptimal for edge devices with limited processing power [12]. Second, while multimodal approaches are effective, they limit flexibility in real-world scenarios where only audio is available. Third, most diffusion methods rely on complex architectures such as large-scale U-Nets, which are unsuitable for low-latency applications [25], [26].

On the other hand, although neural codecs like SoundStream and EnCodec have demonstrated efficiency in audio compression, the resulting latent representations have been rarely explored for diffusion-based restoration and denoising tasks. Operating in the latent space has the potential to reduce data dimensionality by up to 90%, enabling faster inference without compromising reconstruction quality [7], [59].

To address these limitations, this study proposes WaveLLDM (Wave Lightweight Latent Diffusion Models), an end-to-end architecture that combines a neural audio codec with a lightweight diffusion model operating in a compressed latent space [13], [20]. By functioning entirely in the latent domain, WaveLLDM reduces computational load while preserving audio quality, leveraging ConvNeXt blocks and rotational attention mechanisms to maintain temporal context [27], [50], [52], [56].

Training is conducted in two phases: pre-training for compression with spectral regularization, and restoration via diffusion to reconstruct missing segments [9]. The architecture is capable of recovering speech within a 50–450 ms range without additional modalities, while supporting real-time processing on edge devices. By unifying compression, restoration, and denoising, WaveLLDM addresses the fragmentation and inefficiency of earlier approaches.

## II. RELATED WORKS

Numerous studies have explored a wide range of architectures and frameworks for speech enhancement and restoration tasks. One such study introduced MP-SENet, a neural network designed to perform magnitude and phase spectral denoising in parallel [28]. The MP-SENet codec architecture integrates a convolution-augmented transformer-based encoder-decoder with two separate decoding paths for magnitude masking and phase estimation. Training involves a multi-level loss computed over magnitude spectra, phase spectra, short-time complex spectra, and time-domain signals. As a result, MP-SENet achieved a PESQ score of 3.50 and STOI of 0.96 on the VoiceBank+DEMAND dataset.

Another study investigated generative approaches for speech enhancement by comparing score-based models with the Schrödinger Bridge (SB) framework [40]. Through comprehensive experiments, the authors highlighted the differences in training behavior and performance between these two approaches. Furthermore, they proposed a novel perceptual loss function tailored to the SB framework, which significantly improved the perceptual quality of speech signals, although the STOI scores remained around 0.86.

To address the limitation of diffusion models that assume a single Gaussian noise distribution, Diffusion Gaussian Mixture Audio Denoise (DiffGMM) was proposed [54]. By incorporating Gaussian Mixture Models (GMMs) into the reverse diffusion process, DiffGMM employs a 1D U-Net for feature extraction and estimates the GMM parameters using a linear layer. This approach effectively handles more complex noise types and achieves state-of-the-art performance with PESQ of 3.48 and STOI of 0.96.

Another notable contribution is Mamba-SEUNet, a hybrid architecture that combines the Mamba state-space model with a classical U-Net structure [53]. Mamba-SEUNet models forward and backward signal dependencies at multiple resolutions and leverages skip connections to capture multi-scale information. On the VCTK+DEMAND dataset, the model achieved a PESQ score of 3.59 with low computational complexity, and further improvements were observed (PESQ up to 3.73) after integrating a Perceptual Contrast Stretching technique.

## III. RESEARCH METHODOLOGY

WaveLLDM is a two-stage model that integrates FireflyGAN as a neural codec and a Latent Diffusion Model (LDM) for efficient audio restoration and denoising in the latent space. As illustrated in Figures 3.1 and 3.2, FireflyGAN consists of a ConvNeXt encoder, GFSQ quantization, and a HiFi-GAN decoder [24], [31], [39]. The ConvNeXt architecture was selected due to its superior performance over conventional CNNs, as demonstrated in ImageNet evaluations [27], [56], while GFSQ offers more efficient codebook utilization compared to other vector quantization approaches [31].

The HiFi-GAN decoder is simplified by replacing the original multi-receptive field (MRF) module with a Parallel Block, enabling efficient mapping from latent representations to audio waveforms [19], [24], [48]. The ConvNeXt encoder employs a staged arrangement of ConvNeXt blocks, 1D convolutions, and layer normalization to extract high-level features. Each ConvNeXt block consists of depthwise convolution, pointwise convolution, GELU activation, gamma scaling, drop path, and residual connections. The HiFi-GAN decoder includes upsampling blocks based on SiLU activations, transposed convolutions, and a final Tanh activation.

The resulting latent features are then processed by a probabilistic diffusion model for denoising, implemented as a Denoising Diffusion Probabilistic Model (DDPM) based on a lightweight U-Net architecture. This U-Net leverages temporal ConvNeXt blocks and rotational attention mechanisms to effectively model the temporal dynamics of speech signals [14], [50].

### A. FireflyGAN as a Neural Audio Codec

The WaveLLDM neural audio codec operates as follows. A batch of raw audio waveforms $x \in \mathbb{R}^{B \times 1 \times N}$, consisting of $B$ samples each with $N$ mono waveform points, is transformed into $X \in \mathbb{R}^{B \times n_{mel} \times L}$, where $n_{mel}$ denotes the number of Mel bins. In this setup, $n_{mel} = 160$.

**Algorithm 1** *Training*

1. **repeat**
2.     $x_0 \sim q(x)$
3.     $t \sim \text{Uniform}(\{1, \dots, T\})$
4.     $\epsilon \sim \mathcal{N}(\mathbf{0}, \mathbf{I})$
5.     Apply one step optimization with
6.       $\nabla_\theta \mathbb{E}_{x_0, \epsilon_t} \|\epsilon_t - \epsilon_\theta(x_t, t)\|^2$
7. **until** converges

**Algorithm 2** *Sampling*

1. $x_T \sim \mathcal{N}(\mathbf{0}, \mathbf{I})$
2. **for** $t = T, \dots, 1$ **do**
3.     $z \sim \mathcal{N}(\mathbf{0}, \mathbf{I})$ if $t > 1$ else $z = \mathbf{0}$
4.     $x_{t-1} = \frac{1}{\sqrt{\alpha_t}} \left( x_t - \frac{1-\alpha_t}{\sqrt{1-\bar{\alpha}_t}} \epsilon_\theta(x_t, t) \right) + \sigma_t z$
5. **end for**
6. **return** $x_0$

This spectrogram is then processed by an encoder $\mathcal{E}: \mathbb{R}^{B \times n_{mel} \times L} \to \mathbb{R}^{B \times d \times L}$, yielding the latent representation:

$$z = \mathcal{E}(X), \quad (1)$$

Where $z \in \mathbb{R}^{B \times d \times L}$. To reduce dimensionality, the latent vector is quantized into a discrete representation using Grouped Finite Scalar Quantization (GFSQ) [10], [31]. The process begins by downsampling $z$:

$$z_{down} = f_{down}(z), \quad (2)$$

resulting in $z_{down} \in \mathbb{R}^{B \times C_{down} \times L_{down}}$. GFSQ is controlled by two hyperparameters: the number of codebook channels, $c$, the number of levels per channel, $\mathcal{L} = [L_1, L_2, \dots, L_c]$. For instance, a level configuration of $\mathcal{L} = [8,5,5,5]$ implies $c = 4$. The original channel dimension $C_{down}$ is projected to match $c$ using a neural network layer aligned with the FSQ design, resulting in $z_{down} \in \mathbb{R}^{B \times c \times L_{down}}$. Next, $z_{down}$ is partitioned into $G$-groups:

$$z_{down, grouped} = \left\{ z_b^{(1)}, z_b^{(2)}, z_b^{(3)}, \dots, z_b^{(G)} \right\}_{b=1}^{B} \quad (3)$$

Each scalar element, $z_{b,c,l}^{(g)}$, in a group is quantized using the function:

$$Q: z \to \text{round}(\lfloor L_c/2 \rfloor \cdot \tanh(z)), c \in \{1, 2, \dots, c\}, \quad (4)$$

producing quantized values:

$$\hat{z}_{b,c,l}^{(g)} = Q\left(z_{b,c,l}^{(g)}\right). \quad (5)$$

Each scalar is then mapped to a codebook index $k_{b,c,l}^{(g)}$ allowing decoding via:

$$\hat{z}_{b,c,l}^{(g)} = \text{Codebook}^{(g)}\left[k_{b,c,l}^{(g)}\right]. \quad (6)$$

All vectors from each group are concatenated to form:

$$z_{q,\text{down}} = \left\{ \text{Concat}_{g=1}^{G}\left(z_b^{(g)}\right) \middle| b = 1, \dots, B \right\}. \quad (7)$$

An upsampling function is then applied to produce the final quantized latent vector $z_q$:

$$z_q = f_{up}(z_{q,down}). \quad (8)$$

The reconstructed audio $\hat{x}$ is generated using the Firefly-HiFiGAN decoder, $\mathcal{D}$:

$$\hat{x} = \mathcal{D}(z_q), \quad (9)$$

which represents the reconstructed version of the original waveform $x$. Finally, the diffusion process is applied in the latent space using $z_{down}$. This representation is preferred over $z_q$ for diffusion because it retains continuous latent information, making it more suitable for probabilistic modeling and denoising tasks [31].

To train FireflyGAN, this study adopts adversarial training, supported by several loss functions to guide the learning process. First, the Multi-Scale Mel-Spectrogram Loss is used to measure the reconstruction error between the original and reconstructed signals in a frequency domain aligned with human auditory perception. It is defined as the L1 norm (mean absolute error) between the original Mel-spectrogram $S_s$ and its reconstruction ($\widehat{S_s}$):

$$\mathcal{L}_s = \frac{1}{T \cdot F} \sum_{t=1}^{T} \sum_{f=1}^{F} \left| S_s(t, f) - \widehat{S_s}(t, f) \right|, \quad (10)$$

where $T$ is the number of time frames, $F$ is the number of Mel frequency bins, and $S_s(t, f)$ denotes the spectral energy at time frame $t$ and frequency bin $f$ for a given scale $s$ [49].

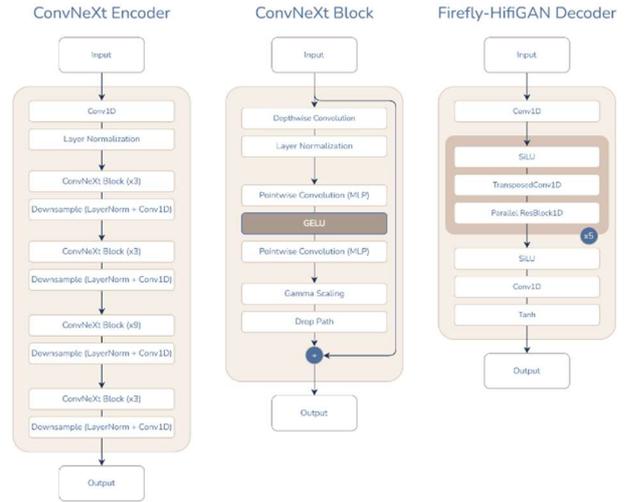

**Figure 3.1** FireflyGAN Components

Second, the Multi-Scale Spectral Loss measures the discrepancy between the target audio signal $x$ and the reconstructed signal $\hat{x}$ in the frequency domain. This is computed by accumulating the L1 distances between magnitude spectrograms obtained from the Short-Time Fourier Transform (STFT) at multiple resolutions [38]. Formally, the loss is defined as:

$$\mathcal{L}_{spectral} = \sum_{i=1}^{N} \left\| \left| \text{STFT}\left(x; n_{fft}^{(i)}, win_{length}^{(i)}, hop_{length}^{(i)}\right) \right| - \left| \text{STFT}\left(\hat{x}; n_{fft}^{(i)}, win_{length}^{(i)}, hop_{length}^{(i)}\right) \right| \right\|_1, \quad (11)$$

which is a collection of different parameter configurations. This approach integrates spectral information at various resolutions, thus improving frequency representation and overall audio reconstruction quality. Third, the loss $\left\{ \left( n_{fft}^{(i)}, win_{leng}^{(i)}, hop_{lengt}^{(i)} \right) \right\}_{i=1}^{N}$ feature matching function is used to minimize the difference between the high-level features extracted from the original data and the generated result data on the discriminator during adversarial training [19]. The basic principle is to force the distribution of the generated data feature ($\hat{x}$) to merge with the distribution of the original data feature ($x$) in an information-rich representation space. Mathematically, for the to- layer of the feature extraction network ($\Phi_l$), the loss function is calculated as the L1 or L2 norm of the difference in features of the two distributions:

$$\mathcal{L}_{feat}^{(l)} = \frac{1}{C_l \cdot H_l \cdot W_l} \sum_{c=1}^{C_l} \sum_{h=1}^{H_l} \sum_{w=1}^{W_l} \|\Phi_l(x)[c,h,w] - \Phi_l(\hat{x})[c,h,w]\|_1, \quad (12)$$

where $C_l, H_l$, and $W_l$ each are the channel dimensions, height, and width of the feature matrix on the $l$-th layer. This approach assumes that similarities in neural network feature spaces reflect perceptual similarities, so that the generation results are not only pixelally accurate but also semantically accurate. In multilevel implementations, the losses of multiple layers are weighted together to capture feature abstractions at different scales:

$$\mathcal{L}_{fm} = \sum_{l=1}^{L} \lambda_l \cdot \mathcal{L}_{feat}^{(l)}, \quad (13)$$

with $\lambda_l$ as the coefficient of weight for the $l$-th layer. The initial layer tends to emphasize texture and edges (low-level features), while the final layer focuses on global structures and objects (high-level features).

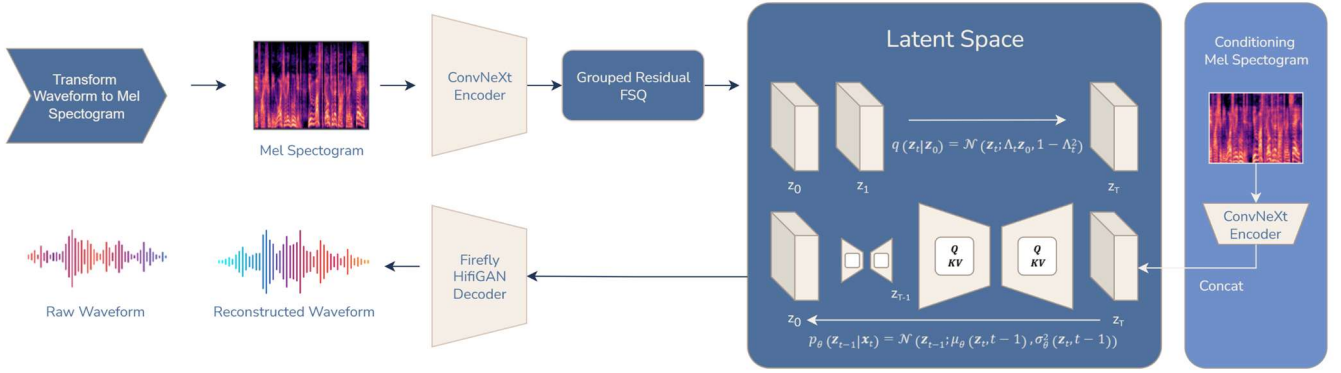

**Figure 3.2** WaveLLDM Architecture

*B. Latent Denoising Diffusion Probabilistic Model*

This study adopts the diffusion model paradigm as *a noise estimator* instead of a score-based estimator. This is because this paradigm is relatively easier, simpler, and able to provide state-of-the-art performance [32], [36], [47], [58], [60]. In the process, the trained Neural audio codecs transform $X$ into latent representation $z_0 = z_{down} \in \mathbb{R}^{B \times C_{down} \times L_{down}}$. A pair of latent representations $z_0$ and $z_0'$ are defined, where $z_{down}$ is a latent representation extracted from clean speech voice audio and $z_{down}'$ is a latent representation extracted from degraded speech voice audio. An estimator $\epsilon_\theta$ learns to predict the Gaussian noise added to $z_0$ through time steps$t$, guided by a variance scheduler $\{\beta_t \in (0,1)\}_{t=1}^T$ that controls the noise intensity [9], [41], [44]. This process, known as forward diffusion, follows the formulation:

$$z_t = \sqrt{\alpha_t} z_{t-1} + \sqrt{1-\alpha_t} \epsilon_{t-1}, \epsilon \in \mathcal{N}(0, I), \quad (14)$$

where $\alpha_t = 1 - \beta_t$. Defining the cumulative product $\overline{\alpha_t} = \prod_{i=1}^{t} \alpha_i$, equation (14) simplifies to:

$$z_t = \sqrt{\overline{\alpha_t}} z_0 + \sqrt{1-\overline{\alpha_t}} \epsilon. \quad (15)$$

In general, larger update steps are applied when the sample becomes noisier, so $\beta_1 < \beta_2 < \cdots < \beta_T$, which implies $\overline{\alpha_1} > \overline{\alpha_2} > \cdots > \overline{\alpha_T}$.

The reverse diffusion process is defined as a stochastic inverse:

$$z_{t-1} = \frac{1}{\sqrt{\alpha_t}} \left( z_t - \frac{1-\alpha_t}{\sqrt{1-\overline{\alpha_t}}} \epsilon_\theta(z_t, z_0', t) \right) + \sigma_t z, \quad (16)$$

with $\sigma_t = \sqrt{\beta_t}$. The probabilistic estimator is optimized using a simple loss function:

$$\theta^* = \underset{\theta}{\mathrm{argmin}} \, \mathbb{E}_{z_0, \epsilon_t} \|\epsilon_t - \epsilon_\theta(\sqrt{\overline{\alpha_t}} z_0 + \sqrt{1-\overline{\alpha_t}} \epsilon, z_0', t)\|^2. \quad (17)$$

This optimization continues until the gradient $\nabla_\theta \mathbb{E}_{z_0, \epsilon_t} \|\epsilon_t - \epsilon_\theta(\sqrt{\overline{\alpha_t}} z_0 + \sqrt{1-\overline{\alpha_t}} \epsilon, z_0', t)\|^2$ converges [13], [14], [20], [22].

*C. Rotary U-Net as Probabilistic Estimator*

The proposed U-Net architecture (Figure 3.3) processes latent audio representations of shape $(B, d, L)$ where $B$ denotes the batch size, $d$ represents the feature dimension, and $L$ indicates the sequence length. The initial convolutional layer maps the input from $(B, d, L)$ to $(B, c_{base}, d, L)$, which serves as the initial projection of the latent features, with $c_{base} = 64$. The network continues through a downsampling path composed of several stages. Each stage includes a residual block activated by the SiLU function and a Rotary Attention layer. The Temporal ConvNeXt block (T-ConvNeXt) refines features by incorporating timestep embeddings and conditioning through Feature-wise Linear Modulation (FiLM) using the following formulation:

$$\Gamma, B = \mathrm{Conv}_{1 \times 1}(\mathrm{concat}[\delta(t), \psi(z_c)]), \quad \Gamma, B \in \mathbb{R}^{B \times c \times d \times l} \quad (18)$$

$$\tilde{U} = \mathrm{LayerNorm2D}(\mathrm{DWConv}(z_t)), \quad (19)$$

$$\mathrm{FiLM}(\tilde{U}; \Gamma, B) = (1 + \Gamma) \odot \tilde{U} + B, \quad (20)$$

Here, $\delta$ denotes the sinusoidal timestep embedding transformation, and $\psi$ represents the interpolation operator that adjusts the spatial dimensions of $z_c$ to match the feature map $\tilde{U}$ produced by the initial two layers of the ConvNeXt block [37], [41], [56].

The rotary attention mechanism in this architecture combines linear attention with rotational position embedding. Unlike the standard softmax-based multihead attention, linear attention approximates attention computation using a kernel feature map, reducing the overall computational complexity from $O(L^2)$ to $O(L)$ with respect to the sequence length $L$. This approach enables the model to process longer latent audio sequences without excessive memory usage [18].

Formally, let $Q, K$, and $V$ denote query, key, and value matrices obtained from the input feature map $X$. The kernel feature map $\phi(.)$ is applied to $Q$ and $K$ as follows:

$$\tilde{Q} = \phi(Q), \qquad \tilde{K} = \phi(K). \qquad (21)$$

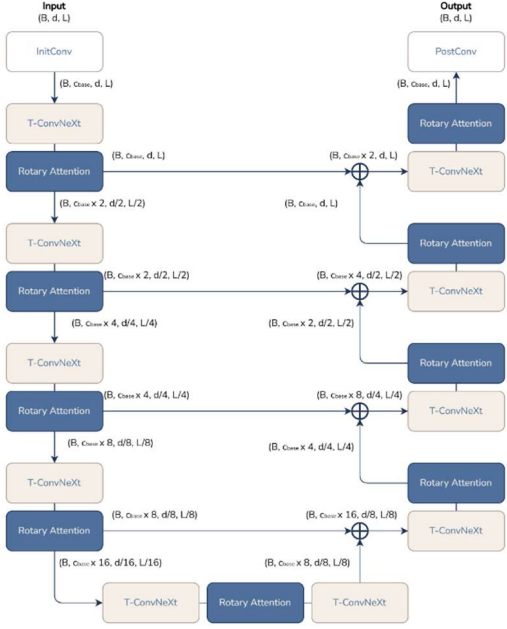

**Figure 3.3** Rotary U-Net architecture

The linear attention output is then computed by:

$$Attention(X) = \frac{\tilde{Q}(\tilde{K}^T V)}{\tilde{Q}(\tilde{K}^T \mathbf{1})}, \qquad (22)$$

where $\mathbf{1}$ is a vector of ones to normalize the attention weights.

To encode positional information, the query and key vectors are added with rotational position embedding (RoPE) before applying the kernel map $\phi(.)$. For each position index $n$, a rotation matrix $R_n$ is constructed and applied to each channel pair in the query and key vectors. Let $q_n$ and $k_n$ denote the query and key at position $n$. The embedding transforms them as follows:

$$\overline{q_n} = R_n q_n, \qquad \overline{k_n} = R_n k_n. \qquad (23)$$

Here, $R_n$ typically introduces sinusoidal rotations dependent on position $n$. This operation preserves the Euclidean norm of $q_n$ and $k_n$ while encoding positional information in a way that supports alignment across the sequence.

Each downsampling operation reduces the spatial dimensions $d$ and $L$ (e.g., $L \to L/2$ dan $d \to d/2$) while increasing the channel dimension $c_{base} \to 2c_{base}$. After several downsampling steps, the bottleneck layer receives a compressed representation of shape $(B, c_{base} \times 16, \frac{d}{16}, \frac{L}{16})$ [12], [14], [25], [26], [41]. This layer follows the same structure, consisting of residual blocks and rotary attention modules, without further reducing the dimensions. The upsampling path mirrors the downsampling stages, using transposed convolutions (or equivalents) to restore the original sequence length. Each upsampling stage is paired with a skip connection from the corresponding downsampling block, merging intermediate features to recover fine-grained information. Finally, the post-convolutional layer maps the decoded features from $(B, c_{base}, d, L)$ back to $(B, d, L)$.

## IV. EXPERIMENTS AND RESULTS

### A. Dataset

This research uses a combined dataset of LibriVox and Voicebank+DEMAND, totaling 143,555 samples with an estimated duration of 160 hours of speech [15], [57]. For data preprocessing, each audio sample is resampled to 48 kHz and normalized in duration to 0.68 seconds for the first training stage and 4.78 seconds for the second stage. To address noise in datasets other than Voicebank+DEMAND, which already contains clean-noisy pairs, the authors collect natural noise samples from online sources and field recordings. Evaluation takes place on the Voicebank+DEMAND test set using three metrics: Perceptual Evaluation of Speech Quality (PESQ), Short-Time Objective Intelligibility (STOI), and Log Spectral Distance (LSD) [9], [25], [26], [44].

**Table 4.1** Comparison of Model Performance Results in Speech Audio Inpainting on the Voicebank+DEMAND Test Set. The ($\downarrow$) symbol indicates that lower metric values correspond to better performance.

| Methods | Params (M) | Mask (ms) | LSD ($\downarrow$) |
|---|---|---|---|
| WaveLLDM-S | 13,98 | 0 | 0,52 |
|  |  | 50 | 0,53 |
|  |  | 250 | 0,53 |
|  |  | 450 | 0,55 |
| WaveLLDM-Base | 51,93 | 0 | **0,48** |
|  |  | 50 | 0,53 |
|  |  | 250 | 0,54 |
|  |  | 450 | 0,60 |
| AudioLDM-S | 181,00 | 0 | 1,12 |
| AudioLDM-L | 739,00 | 0 | 0,98 |

## B. Model Configurations

The training process for WaveLLDM is divided into two stages: training the autoencoder (neural audio codec) and training the diffusion model. The first stage trains the autoencoder to learn latent representations of audio and to reconstruct the original signal from these representations. The encoder produces the latent representation, while the decoder reconstructs the audio from it. The goal of this stage is to preserve high-quality audio information in the latent space. The model is trained for 250,000 steps with a batch size of 16 and 32,768 audio samples per sequence (approximately 0.74 seconds of audio). The Adam optimizer is used with a learning rate of $2 \times 10^{-4}$, along with $\beta_1 = 0.8$, $\beta_2 = 0.99$, and exponential learning rate decay with factor $\gamma = 0.998$ every 2,000 steps. The loss function is a weighted combination of four components, formulated as follows:

$$\mathcal{L}_{total} = \mathcal{L}_{adv} + \lambda_{mel}\mathcal{L}_{mel} + \lambda_{spectral}\mathcal{L}_{spectral} + \lambda_{fm}\mathcal{L}_{fm}, \quad (24)$$

where the weighting coefficients are set as $\lambda_{mel} = 30, \lambda_{spectral} = 20$, and $\lambda_{fm} = 2$ [19]. In addition, the multiscale mel-spectrogram loss assigns an equal weight of $\lambda_s = 1$ for each scale.

In the second training stage, the authors train the Rotary U-Net as a noise estimator in the latent space learned by the neural audio codec in the first stage. This phase optimizes the model to predict noise components in the latent representation using the diffusion process. To ensure stable convergence and high-quality sample generation, the authors apply the following training configuration. The model is trained for 500 epochs ($15.4 \times 10^5$ step) with a batch size of 36 [46]. To maintain stability and efficient parameter updates, the authors use the AdamW optimizer with hyperparameters $\beta_1 = 0.9, \beta_2 = 0.999$, and weight decay ranging between $10^{-2}$ dan $10^{-4}$. The initial learning rate is set to $2 \times 10^{-4}$ and follows exponential decay with factor $\gamma = 0.998$ every 2,500 steps. Additionally, the audio length is limited to 229,376 samples during the second stage of training.

## C. Results

This study trains two variants of the WaveLLDM model, namely WaveLLDM-S and WaveLLDM-Base, with 13.98 million and 51.93 million parameters respectively. To evaluate model performance, the authors use two objective metrics: PESQ (wide-band) and STOI (Short-Time Objective Intelligibility). Both metrics require paired target and reference signals for evaluation and have been widely adopted in research related to speech quality and clarity enhancement. However, it is important to note that these metrics have limitations when assessing outputs from generative models, as parts of the generated signal may not perfectly reconstruct the reference. Nonetheless, PESQ and STOI still provide strong indications of perceptual quality and speech intelligibility, so the authors consider them relevant and sufficiently representative for evaluating the system's performance. The evaluation results are summarized in Table 4.1 and Table 4.2.

**Tabel 4.2** Comparison of Model Performance and Previous State-of-the-Art on the Voicebank+DEMAND Test Set. "–" indicates unavailable information. The (↑) symbol denotes that higher metric values indicate better performance.

| Metods | WB-PESQ (↑) | STOI (↑) | Params (M) |
|---|---|---|---|
| Mamba-SEUNet | 3,59 | **0,96** | 6,28 |
| Schrodinger Bridge | **3,70** | 0,86 | 65,6 |
| DiffGMM | 3,48 | 0,96 | - |
| MP-SENet | 3,50 | 0,96 | **2,05** |
| **WaveLLDM-Small** | 1,62 | 0,76 | 13,98 |
| **WaveLLDM-Base** | 1,71 | 0,78 | 51,93 |

This study also evaluates the speech audio restoration capability using the Log-Spectral Distance (LSD) metric. Table 4.1 presents the performance of both WaveLLDM-Small (S) and WaveLLDM-Base. The performance results, as summarized in Table 4.2, show that the model achieves a WB-PESQ score of 1.62 for the WaveLLDM-Small variant and 1.71 for WaveLLDM-Base, along with STOI scores of 0.76 and 0.78 respectively on the Voicebank+DEMAND dataset. These scores fall significantly below those achieved by state-of-the-art methods such as Mamba-SEUNet (WB-PESQ: 3.59, STOI: 0.96), Schrodinger Bridge (WB-PESQ: 3.70, STOI: 0.86), DiffGMM (WB-PESQ: 3.48, STOI: 0.96), and MP-SENet (WB-PESQ: 3.50, STOI: 0.96). This performance gap indicates that WaveLLDM, in its current configuration, does not yet match the perceptual quality and clarity delivered by these methods. Several factors may contribute to this limitation, such as suboptimal architectural complexity or training constraints, including the absence of a fine-tuning phase and limited training duration. For comparison, Hi-Res LDM (Dhyani et al., 2024) trains a latent diffusion model for audio inpainting on a dataset with a total audio duration of 1,250 hours (approximately 1 million samples). In contrast, WaveLLDM only uses 143,555 samples.

Despite this, WaveLLDM demonstrates a clear architectural advantage by integrating a FireflyGAN-based neural audio codec with a RotaryUNet-based latent diffusion model. This approach establishes a strong foundation for further development, especially for restoration tasks that require long-term dependency modeling and efficient latent representations. Evaluation on the speech audio inpainting task, as shown in Table 4.2, further supports the model's potential. The low Log-Spectral Distance (LSD) values, ranging from 0.48 to 0.60 across both variants, indicate the model's ability to reconstruct audio spectra effectively, even under conditions where segments of the signal are missing (masked), as illustrated in Figure 4.4. These results reflect adequate spectral fidelity, a key indicator of audio quality in restoration contexts.

Additional experiments involving directly spoken samples with natural noise further confirm WaveLLDM's ability to generalize beyond the training dataset. These findings suggest that the model adapts well to real-world environmental conditions, although further testing is necessary to validate its performance consistency.

## V. CONCLUSIONS

This research successfully develops the WaveLLDM model for audio restoration, focusing on speech denoising and speech audio inpainting, and deploys it in an interactive Gradio-based application. The model reconstructs audio spectra effectively, with low Log-Spectral Distance (LSD) values (0.48–0.60) and strong adaptation to unseen data. However, it remains inferior to state-of-the-art methods such as Mamba-SEUNet in terms of perceptual quality and speech clarity, achieving lower WB-PESQ scores (1.62–1.71) and STOI scores (0.76–0.78), compared to WB-PESQ 3.59 and STOI 0.96. This performance gap stems from the lack of architectural optimization, the absence of a fine-tuning phase, and insufficient training duration. Nevertheless, the model's flexible architecture, which combines a neural audio codec with a latent diffusion model, provides a strong foundation for further development.

## ACKNOWLEDGEMENTS

The author expresses deep gratitude to Chieh-Hsin Lai, a researcher at Sony AI Japan specializing in diffusion models for speech processing, for his valuable feedback and insightful perspectives. This research also receives support through access to an NVIDIA DGX-A100 machine provided by the Faculty of Intelligent Electrical and Informatics Technology (FT-EIC), Institut Teknologi Sepuluh Nopember.